\documentclass[sn-mathphys]{sn-jnl}

\jyear{2022}%

\theoremstyle{thmstyleone}%
\theoremstyle{thmstyletwo}%

\theoremstyle{thmstylethree}%

\begin{document}

\title[Speed of sound in QCD matter at finite temperature and density]{Speed of sound in QCD matter at finite temperature and density}


\author*[1]{\fnm{Guo-yun} \sur{Shao}}\email{gyshao@mail.xjtu.edu.cn}

\author[1]{\fnm{Xin-ran} \sur{Yang}}

\author[1]{\fnm{Chong-long} \sur{Xie}}
\author[2]{\fnm{Wei-bo} \sur{He}}


\affil[1]{\orgdiv{School of Physics}, \orgname{ Xi’an Jiaotong University}, \orgaddress{\city{Xi’an}, \postcode{710049}, \country{China}}}
\affil[2]{\orgdiv{School of Physics}, \orgname{ Peking University}, \orgaddress{\city{Beijing}, \postcode{100871},  \country{China}}}


\abstract{The speed of sound in QCD matter at  finite temperature and density is investigated within the Polyakov loop improved Nambu--Jona-Lasinio (PNJL)  model. The spinodal  structure associated with the chiral first-order chiral phase transition is considered 
to describe the continuous variation of the speed of sound. The behaviors of the squared sound speed in different phases, including the stable, metastable and unstable phases, are derived. The relation between speed of sound and QCD phase transitions is systematically explored. In particular, the boundary of vanishing sound velocity is derived in the temperature-density phase diagram, and the region where the sound wave equation being broken is pointed out.  Some interesting features of speed of sound under different definitions are also discussed. }




\maketitle

\section{Introduction}
Quark-gluon plasma~(QGP) can be created in heavy-ion collision~(HIC) experiments at relativistic energies. A crucial topic relevant is to explore the equation of state~(EOS) and phase transition from QGP to hadronic matter. 
The hydrodynamic simulation provides  a method to study the EOS of QGP~\cite{Song11, Song112, Deb16}. During the space-time evolution of QCD matter, the speed of sound is one of the crucial  physical quantities. Its dependence on environment  (temperature, density, chemical potential, etc.) carries important information in describing the evolution of the fireball and final observables.  Recently, the studies in Refs.~\cite{Gardim20,Sahu21,Biswas20} show that the speed of sound  as a function of charged particle multiplicity $\langle d N_{ch}/d\eta \rangle$  can be extracted from heavy-ion collision data. In Ref.~\cite{Sorensen21} the authors try to build a connection between the sound speed and  baryon number cumulants to study the QCD phase structure.

The speed of sound in neutron star has also received a lot of attention~(e.g., Refs.~\cite{Reed20, Kanakis20,Han20}).
The density dependent behavior of sound velocity influences the mass-radius relation, the tidal deformability, and provides a sensitive probe of the EOS of neutron star matter. To obtain a two solar mass neutron star, some studies find that it is essential for neutron star matter to have a density range where the EOS is very stiff and the corresponding squared speed of sound is significantly larger than $1/3$~\cite{Tews18,Greif19,Forbes19,Drischler20,Essick20,Han19,Kojo}. The study in Ref.~\cite{Jaikumar21}  indicates that the speed of sound is crucial for the gravitational wave frequencies induced by the $g$-mode oscillation of a neutron star. 
It is also interesting  to study the gravitational wave induced by the cosmic QCD phase transition in which the speed of sound plays a significant role.

As an important quantum in describing the evolution of strongly interacting matter, the  relation between  the speed of sound  and QCD phase transition is worth exploring. The speed of sound  has been calculated, e.g.,  in 
lattice QCD~\cite{Aoki06, Borsanyi14, Bazavov14, Philipsen13, Borsanyi20},  (P)NJL model~\cite{Motta18,Ghosh06,Marty13,Deb16,Saha18,zhao20}, quark-meson coupling model~\cite{Schaefer10,Abhishek18},  hadron resonance gas (HRG) model~\cite{Venugopalan92,Bluhm14}, field correlator method~(FCM)~\cite{Khaidukov18, Khaidukov19} and quasiparticle model~\cite{Mykhaylova21}.
In previous studies, the main focus is put on the region of high temperature and  vanishing or small chemical potential. 
In Ref.~\cite{he2022}, we give an intensive study on the speed of sound in QCD matter in the full temperature-chemical potential phase diagram. The numerical results indicate that the dependence of sound speed on temperature and chemical potential is indicative of QCD phase transition.  

However, only the sound speed in the stable phase is considered in  Ref.~\cite{he2022}. There are still some crucial issues that need to be clarified. First of all, the spinodal structure may be involved in heavy-ion collision experiments with the decrease of collision energy~\cite{Mishustin1999, Randrup2004, Koch2005, Sasaki2007, Sasaki2008, Randrup2009, Steinheimer2012, Li2016, Steinheimer2016, Steinheimer2017,Shao2020}. A complete evolution of sound speed in the 
metastable and unstable phases needs to be explored to give a distinct description of the fireball expansion. Secondly, it is found that the sound speed takes small values  at the CEP and on the boundaries of the first-order phase transition near the CEP~\cite{he2022}. A question aroused is that where is the boundary of vanishing speed of sound. Furthermore, the behavior of sound speed in the temperature and density phase diagram is still not explored.

On the other hand, the values of speed of sound under different conditions are involved in dealing with different problems in nuclear physics, such as the gravitational signal from cosmic QCD phase transition~\cite{Ahmadvand2018,Tenkanen2022}, the bulk viscosity of strongly interacting matter~\cite{Deb16}, the equation of state of neutron star matter~\cite{Reed20, Kanakis20,Han20} and the evolution of QGP in HIC experiments.  In this work, we will give a systematic study on the relation between speed of sound and QCD phase transitions  at finite temperature and density under several different constraint conditions. This work is helpful in dealing with the physics problems mentioned above. 


The paper is organized as follows. In Sec.~II, we derive the formulae of speed of sound under different definitions in the temperature and density space, and then briefly introduce  the 2+1 flavor PNJL quark model. In Sec.~III, we present the numerical results of squared sound speed and discuss the relations with the QCD phase structure. A summary is finally given in Sec. IV.

\section{Speed of sound and the PNJL quark model}
The general definition of speed of sound is
\begin{equation}
c^2_{X}=\left(\frac{\partial p}{\partial \epsilon}\right)_{X}.
\end{equation}
A specifying constant quantity $X$ is required to describe the propagation of the compression wave through a medium.
To indicate the different profiles of QCD matter, $X$ can be chosen as $s/\rho_B, s, \rho_B, T, \mu_B$.
Different definitions of speed of sound are taken in practice in dealing with different physics issues.

For a fireball created in relativistic heavy-ion collisions, it evolves with a constant entropy density per baryon $s/\rho_B$ if it is taken as an ideal fluid. 
Therefore, it is  meaningful to calculate the speed of sound along the isentropic curve 
\begin{equation} 
\label{cs}
c_{s/{\rho_B}}^2=\bigg(\frac{\partial p}{\partial \epsilon}\bigg)_{s/{\rho_B}}.
\end{equation}
The dependence of $c_{s/{\rho_B}}^2$ on parameters, e.g., temperature and density, can indicate the variation of sound speed during the evolution and provide important knowledge of interaction, phase transition and  the EOS of QGP. 


The speed of sound with constant baryon number density or entropy density are taken in describing the intermediate process of a hydrodynamic evolution~\cite{Deb16},
\begin{equation}
\label{}
c_{\rho_B}^2=\bigg(\frac{\partial p}{\partial \epsilon}\bigg)_{\rho_B} \quad \textrm{and} \quad c_s^2=\bigg(\frac{\partial p}{\partial \epsilon}\bigg)_s.
\end{equation}
For example, the temporal derivatives of temperature and  chemical potential are functions of  $c_{\rho_B}^2$ and $c_s^2$, as
\begin{equation}
\label{4}
\partial_0 \mu_B=-c_{s}^2  \mu_B\, \bold{\nabla}\cdot \bold{u},    
\end{equation}
and 
\begin{equation}
\label{5}
\partial_0 T=-c_{\rho_B}^2  T\, \bold{\nabla}\cdot \bold{u},    
\end{equation}
where $\bold u$ denotes the space component of four-velocity. The values of $c_{\rho_B}^2$ and $c_s^2$ are directly connected to the bulk viscosity coefficient. 

It is also interesting to calculate the sound speed with a fixed temperature or chemical potential
\begin{equation}
\label{ }
c_T^2=\bigg(\frac{\partial p}{\partial \epsilon}\bigg)_T, \,\,\,\,\,\,\,\,c_{\mu_B}^2=\bigg(\frac{\partial p}{\partial \epsilon}\bigg)_{\mu_B},
\end{equation}
In Ref.~\cite{Sorensen21} the authors  estimate $c^2_T$ as a function of the logarithmic derivative with respect to the baryon  density of QCD matter, and try to build a connection with the baryon number cumulants to aid in detecting the QCD critical endpoint.  Besides, $c^2_T$ is also usually taken to study the speed of sound in neutron star matter.

In this study, we will explore the speed of sound under different definitions in the full temperature-density space. Since the general definitions can only be used to calculate the sound speed on special trajectories, it is necessary to derive the corresponding formulae in terms of $T$ and $\rho_B$. With the fundamental thermodynamic relations, the  sound speed formulae under different constraint conditions can be
derived as

\begin{equation}\label{eqss1}
\!c^2_{s/\rho_B}\!=\!\frac{\!s^{2}\!+\!\rho_{B}^{2}\!\left[\!\left(\frac{\!\partial \mu_{B}}{\!\partial \rho_{B}}\!\right)_{\!T}\!\left(\frac{\partial s}{\!\partial T}\!\right)_{\!\rho_{\!B}}\!-\!\left(\frac{\partial \mu_{\!B}}{\partial T}\!\right)_{\!\rho_{\!B}}\!\left(\frac{\partial s}{\partial \rho_{\!B}}\!\right)_{\!T}\!\right]\!+\!s \rho_{\!B}\!\left[\!\left(\frac{\partial \mu_{\!B}}{\partial T}\!\right)_{\!\rho_{\!B}}\!-\!\left(\frac{\partial s}{\partial \rho_{\!B}}\!\right)_{\!T}\right]}{\left(T s+\mu_{B} \rho_{B}\right)\left(\frac{\partial s}{\partial T}\right)_{\rho_{B}}},
\end{equation}
\begin{equation}
c_{s}^{2}=\frac{\rho_{B}\left[\left(\frac{\partial s}{\partial T}\right)_{\rho_{B}}\left(\frac{\partial \mu_{B}}{\partial \rho_{B}}\right)_{T}-\left(\frac{\partial s}{\partial \rho_{B}}\right)_{T}\left(\frac{\partial \mu_{B}}{\partial T}\right)_{\rho_{B}}\right]-s\left(\frac{\partial s}{\partial \rho_{B}}\right)_{T}}{\mu_{B}\left(\frac{\partial s}{\partial T}\right)_{\rho_{B}}},
\end{equation}
\begin{equation}
c_{\rho_{B}}^{2}=\frac{s+\rho_{B}\left(\frac{\partial \mu_{B}}{\partial T}\right)_{\rho_{B}}}{T\left(\frac{\partial s}{\partial T}\right)_{\rho_{B}}},\quad \quad \quad c_{T}^{2}=\frac{\rho_{B}\left(\frac{\partial \mu_{B}}{\partial \rho_{B}}\right)_{T}}{T\left(\frac{\partial s}{\partial \rho_{B}}\right)_{T}+\mu_{B}},
\end{equation}
and
\begin{equation}\label{eqss4}
c_{\mu_{B}}^{2}=\frac{s\left(\frac{\partial \mu_{B}}{\partial \rho_{B}}\right)_{T}}{T\left[\left(\frac{\partial s}{\partial T}\right)_{\rho_{B}}\left(\frac{\partial \mu_{B}}{\partial \rho_{B}}\right)_{T}-\left(\frac{\partial \mu_{B}}{\partial T}\right)_{\rho_{B}}\left(\frac{\partial s}{\partial \rho_{B}}\right)_{T}\right]-\mu_{B}\left(\frac{\partial \mu_{B}}{\partial T}\right)_{\rho_{B}}}.
\end{equation}
The details for deriving these formulae  are affiliated in the appendix~\ref{sec:appd}.  The above formulae are only correct for isospin symmetric matter. The corresponding formulae will be much more complicated for isospin asymmetric matter. 

To demonstrate the relation between the speed of sound under different definition and QCD phase structure, we take the 2+1 flavor PNJL quark model in the calculation. 
The Lagrangian density is given by
\begin{eqnarray}
\mathcal{L}&\!=&\!\bar{q}(i\gamma^{\mu}D_{\mu}\!+\!\gamma_0\hat{\mu}\!-\!\hat{m}_{0})q\!+\!
G\sum_{k=0}^{8}\big[(\bar{q}\lambda_{k}q)^{2}\!+\!
(\bar{q}i\gamma_{5}\lambda_{k}q)^{2}\big]\nonumber \\
           &&-K\big[\texttt{det}_{f}(\bar{q}(1+\gamma_{5})q)+\texttt{det}_{f}
(\bar{q}(1-\gamma_{5})q)\big]\nonumber \\ \nonumber \\
&&-U(\Phi[A],\bar{\Phi}[A],T),
\end{eqnarray}
where $q$ denotes the quark fields with three flavors, $u,\ d$, and
$s$; $\hat{m}_{0}=\texttt{diag}(m_{u},\ m_{d},\
m_{s})$ in flavor space; $G$ and $K$ are the four-point and
six-point interacting constants, respectively.  The $\hat{\mu}=diag(\mu_u,\mu_d,\mu_s)$ are the quark chemical potentials.

The covariant derivative in the Lagrangian is defined as $D_\mu=\partial_\mu-iA_\mu$.
The gluon background field $A_\mu=\delta_\mu^0A_0$ is supposed to be homogeneous
and static, with  $A_0=g\mathcal{A}_0^\alpha \frac{\lambda^\alpha}{2}$, where
$\frac{\lambda^\alpha}{2}$ is $SU(3)$ color generators.
The effective potential $U(\Phi[A],\bar{\Phi}[A],T)$ is expressed with the traced Polyakov loop
$\Phi=(\mathrm{Tr}_c L)/N_C$ and its conjugate
$\bar{\Phi}=(\mathrm{Tr}_c L^\dag)/N_C$. The Polyakov loop $L$  is a matrix in color space
\begin{equation}
   L(\vec{x})=\mathcal{P} exp\bigg[i\int_0^\beta d\tau A_4 (\vec{x},\tau)   \bigg],
\end{equation}
where $\beta=1/T$ is the inverse of temperature and $A_4=iA_0$.

The Polyakov-loop effective potential  is
%
\begin{eqnarray}
     \frac{U(\Phi,\bar{\Phi},T)}{T^4}&=&-\frac{a(T)}{2}\bar{\Phi}\Phi +b(T)\mathrm{ln}\big[1-6\bar{\Phi}\Phi+4(\bar{\Phi}^3+\Phi^3)-3(\bar{\Phi}\Phi)^2\big],
\end{eqnarray}
where
\begin{equation}
   \!a(T)\!=\!a_0\!+\!a_1\big(\frac{T_0}{T}\big)\!+\!a_2\big(\frac{T_0}{T}\big)^2 \,\,\,\texttt{and}\,\,\,\,\, b(T)\!=\!b_3\big(\frac{T_0}{T}\big)^3.
\end{equation}
The parameters $a_i$, $b_i$ listed in Table. \ref{tab:1} are fitted according to the lattice simulation of  QCD thermodynamics in
pure gauge sector. 
The $T_0=210$\, MeV
is implemented in the calculation.  
\begin{table}[ht]
\centering
\caption{Parameters in the Polyakov-loop potential~\cite{Robner07}}
\label{tab:1}
\begin{tabular*}{\columnwidth}{@{\extracolsep{\fill}}llll@{}}
\hline
\multicolumn{1}{@{}l}{$a_0$} & $a_1$ & $a_2$ & $b_3$\\
\hline
 $ 3.51$                   & -2.47        &  15.2      & -1.75               \\ 
 \hline
\end{tabular*}
\end{table}

The constituent quark mass in the mean field approximation can be derived as
\begin{equation}
M_{i}=m_{i}-4G\phi_i+2K\phi_j\phi_k\ \ \ \ \ \ (i\neq j\neq k),
\label{mass}
\end{equation}
 where $\phi_i$ stands for quark condensate of the flavor $i$.

The thermodynamical potential of bulk quark matter is derived as
\begin{eqnarray}
\Omega&=&-2T \sum_{i=u,d,s}\int \frac{\mathrm{d}^{3}p}{(2\pi)^{3}} (\mathcal{Q}_1+\mathcal{Q}_2)-2\int_\Lambda \frac{\mathrm{d}^{3}p}{(2\pi)^{3}}3(E_u+E_d+E_s)\nonumber\\
&&+2G\left({\phi_{u}}^{2}+{\phi_{d}}^{2}+{\phi_{s}}^{2}\right)-4K\phi_{u}\,\phi_{d}\,\phi_{s} +U(\bar{\Phi}, \Phi, T)
\end{eqnarray}
where
$\mathcal{Q}_1=\mathrm{ln}(1
+3\Phi e^{-(E_i-\mu_i)/T}+3\bar{\Phi} e^{-2(E_i-\mu_i)/T}+e^{-3(E_i-\mu_i)/T})$,  $\mathcal{Q}_2=\mathrm{ln}(1+3\bar{\Phi} e^{-(E_i+\mu_i)/T}
+3\Phi e^{-2(E_i+\mu_i)/T}+e^{-3(E_i+\mu_i)/T}) $, and
 $E_i=\sqrt{\vec{p}^{\,2}+M_i^2}$ is the dispersion relation.  $\mu_i=\mu_B/3$ is taken for $u,d,s$ quark flavors. 
The pressure $p$ and energy density $\epsilon$ can be derived using the thermodynamic relations in the grand canonical ensemble as
\begin{equation}
\label{ }
P=-\Omega,\,\,\,\,\,\,\,\, \epsilon=-P+Ts+\sum\mu_i \rho_i,
\end{equation}
where $s$ is the entropy density and $\rho_i$ the quark number density of flavor $i$.

For given $T$ and baryon density $\rho_B$, the values of $\phi_{u}, \phi_{d}, \phi_{s}, \Phi$, $\bar{\Phi}$ and $\mu_B$ are determined by solving the equations by minimizing the thermodynamical potential
\begin{equation}
\frac{\partial \Omega^{}}{\partial \phi_{u}}=\frac{\partial \Omega^{}}{\partial \phi_{d}}=\frac{\partial \Omega^{}}{\partial \phi_{s}}=\frac{\partial \Omega^{}}{\partial \Phi}=\frac{\partial \Omega^{}}{\partial \bar{\Phi}}=0,
\end{equation}
and the relevant constraint condition. Other physical quantities can be then derived using thermodynamic relations.  The numerical results of speed of sound under different conditions can be then derived according to Eqs.~(\ref{eqss1})-(\ref{eqss4}).

In the numerical calculation, a cut-off $\Lambda$ is implemented in 3-momentum
space for divergent integrations. We take the model parameters obtained in~\cite{Rehberg96}:
$\Lambda=602.3$ MeV, $G\Lambda^{2}=1.835$, $K\Lambda^{5}=12.36$,
$m_{u,d}=5.5$  and $m_{s}=140.7$ MeV, determined
by fitting $f_{\pi}=92.4$ MeV,  $M_{\pi}=135.0$ MeV, $m_{K}=497.7$ MeV and $m_{\eta}=957.8$ MeV.



\section{Numerical results and discussions }

In this section, we present the numerical results of the speed of sound under different constraint conditions and discuss the relations with the QCD phase transitions.

\subsection {Sound velocity at constant $s/\rho_B$}

\begin{figure} [htbp]
\begin{minipage}{\columnwidth}
\centering
\includegraphics[scale=0.33]{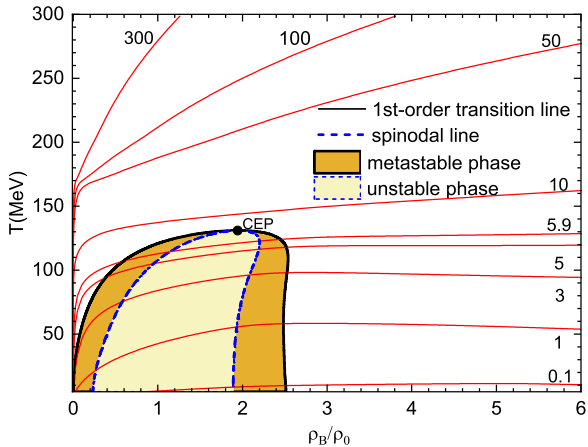}
\end{minipage}
\caption{  QCD phase diagram and the isentropic curves for  $s/\rho_B=0.1, 1, 3, 5, 5.9, 10, 50, 100, 300$  in $T-\rho_B$ plane.}
\label{fig:1}
\end{figure}

Firstly, we plot the QCD phase diagram, including the first-order phase transition~(black solid line) and the spinodal structure~(blue dashed line), which separate the phase diagram into the stable, metastable and unstable phases. 
The first-order phase transition line is obtained according to the thermodynamic conditions for two-phase equilibrium, i.e., $T_1=T_2$, $\mu_{1}=\mu_2$, $P_1=P_2$ for two stable phases. The spinodal line is derived with the mechanical unstable condition. The corresponding inflection points of pressure as a function of density can be determined  for a given $T$.  For more details to derive the phase boundaries, one can refer to Refs.~\cite{Costa10, Shao2018}.
The spinodal phase decomposition  plays a dominant role in the experimental exploration of the first-order nuclear liquid-gas transition\cite{Chomaz2004,Shao20202}. It has inspired the anticipation to identify the first-order chiral transition in high-energy heavy-ion collisions through the spinodal phase separation~\cite{Mishustin1999, Randrup2004, Koch2005, Sasaki2007, Sasaki2008, Randrup2009, Steinheimer2012, Li2016, Steinheimer2016, Steinheimer2017}. The recent
simulation suggests that the spinodal instability can  be triggered within a certain energy range \cite{Steinheimer2016, Steinheimer2017} .  
We also demonstrate in Fig.~\ref{fig:1} the isentropic curves with $s/\rho_B=0.1, 1, 3, 5, 5.9, 10, 50, 100, 300$ in the $T-\rho_B$ plane to indicate the evolutionary trajectories of an ideal fluid at different collision energies.

\begin{figure} [htbp]
\begin{minipage}{\columnwidth}
\centering
\includegraphics[scale=0.33]{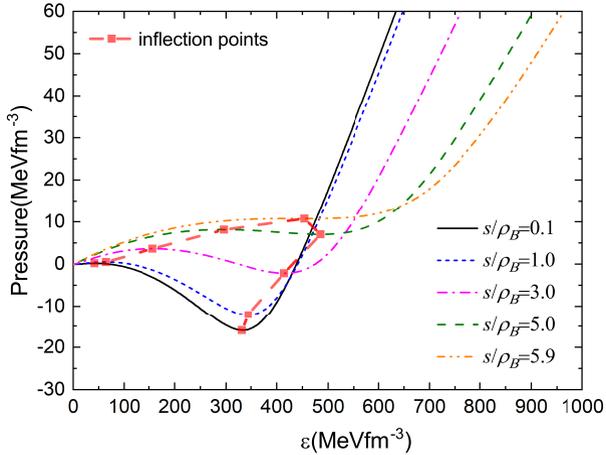}
\end{minipage}
\caption{  Equations of state of QCD matter for $s/\rho_B=0.1, 1, 3, 5, 5.9$. The squares on each curve are the inflection points where $(\frac{\partial p }{\partial \epsilon})_{s/\rho_B}$ changes the sign. The red dashed line is the profile of these inflection points.}
\label{fig:2}
\end{figure}

We present in Fig.~\ref{fig:2} the equations of state with chosen parameter $s/\rho_B=0.1, 1, 3, 5, 5.9$ that pass through the first-order phase transition. For each curve with $s/\rho_B<5.9$, there are two inflection points where $\partial p /\partial \epsilon$ changes the sign. 
The red dashed curve in Fig.~\ref{fig:2} is the connections of these inflection points, which lies in the spinodal 
boundary associated with the first-order phase transition, i.e., in the interior of the unstable phase. We will present the relation clearly in the contour map of speed of sound in the $T-\rho_B$ panel soon.  The inflection points also correspond to the locations  where the sound velocity vanishes in the phase diagram.

\begin{figure} [htbp]
\begin{minipage}{\columnwidth}
\centering
\includegraphics[scale=0.33]{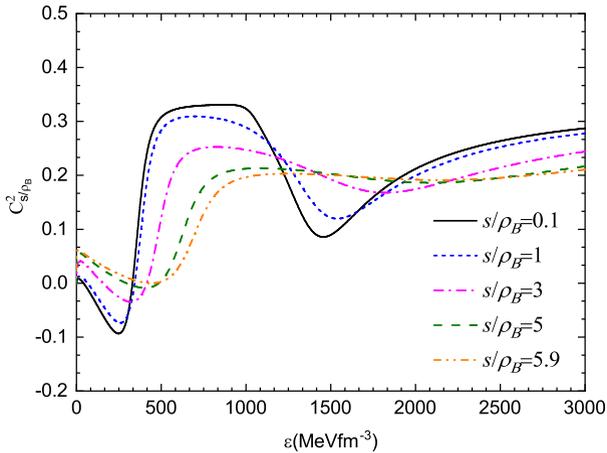}
\end{minipage}
\caption{ Values of $c^2_{s/\rho_B}$ as functions of energy density for  $s/\rho_B=0.1, 1, 3, 5, 5.9$.}
\label{fig:3}
\end{figure}

The square of speed of sound can be directly derived with the definition given in Eq.~(\ref{cs}). In Fig.~\ref{fig:3}, we plot the curve of squared sound speed $c^2_{s/\rho_B}$ as functions of energy density along the evolutionary trajectories for $s/\rho_B=0.1, 1, 3, 5, 5.9$.  Figure~\ref{fig:3} shows that there exists one peak and two valleys on each curve, which indicates that the speed of sound  are closely related to temperature and density. In particular, for the case of $c^2_{s/\rho_B}<0$ in Fig.~\ref{fig:3}, it  corresponds to $\partial p /\partial \epsilon<0$ for a fixed  $s/\rho_B$, as shown in Fig.~\ref{fig:2}.

\begin{figure} [htbp]
\begin{minipage}{\columnwidth}
\centering
\includegraphics[scale=0.3]{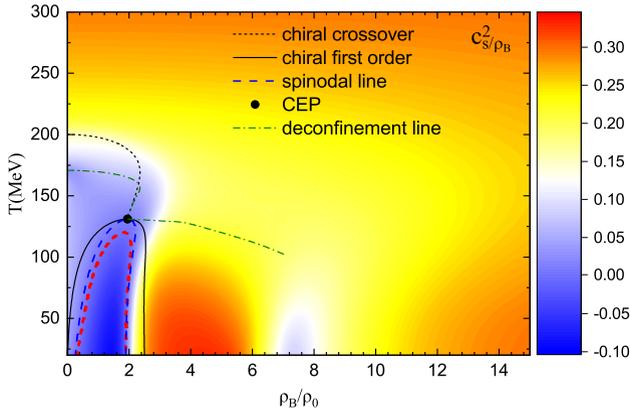}
\end{minipage}
\caption{ Contour map of $c^2_{s/\rho_B}$  in the $T-\rho_B$ plane. The red dashed line is the boundary of vanishing sound velocity .}
\label{fig:4}
\end{figure}

\begin{figure} [htbp]
\begin{minipage}{\columnwidth}
\centering
\includegraphics[scale=0.35]{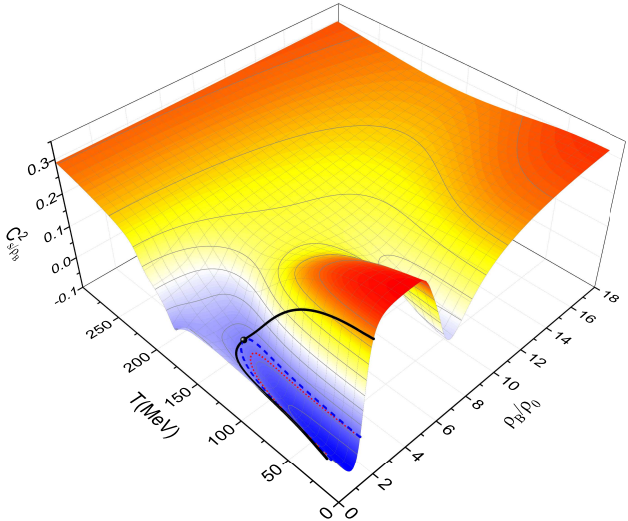}
\end{minipage}
\caption{ 3D map of $c^2_{s/\rho_B}$ as functions of tempeature and density. The black solid line is the first-order chiral phase transition line. The blue dashed line is the spinodal line. The red dashed line is the boundary of vanishing sound velocity.}
\label{fig:5}
\end{figure}

To show more clearly the relation between the speed of sound and QCD phase transitions, we present the contour map and 3D map of $c^2_{s/\rho_B}$ as functions of $T$ and $\rho_B$ in Fig.~\ref{fig:4} and Fig.~\ref{fig:5}, respectively. The phase structure including the chiral crossover, chiral first-order, spinodal and deconfinement phase transitions. The chiral crossover line and deconfinement line are determined by requiring $\partial \phi_i/\partial T$ and $\partial \Phi_i/\partial T$ taking extreme values  for a given chemical potential.

The two figures indicate  that the region around the peaks of $c^2_{s/\rho_B}$ in Fig.~\ref{fig:3} is located in the  region where the chrial symmetry of $u, d$ quark is approximately restored already
but still confined. The valley at high energy density side in Fig.~\ref{fig:3} lies in the region where the chiral condensate of strange quark changes quickly. After the chiral restoration of strange quark the sound speed increases again towards high density. The valley in the low  density side is located in the spinodal region of the first-order phase transition, which is closely related to the chiral condensate of $u, d$ quark. For each value of ${s/\rho_B}$ smaller than 5.9, there exist a range of $(\frac{\partial p }{\partial \epsilon})_{s/\rho_B}$ taking negative values. The red dashed line is the boundary of vanishing sound velocity with $c^2_{s/\rho_B}=0$ derived with Eq.~(\ref{eqss1}) .


Figure~\ref{fig:4} and \ref{fig:5} also show that, at the high temperature or very high density, with the restoration of chiral symmetry $c^2_{s/\rho_B}$ approaches to $1/3$, the value of noninteracting gas. 
The value of $c^2_{s/\rho_B}$ at lower density descends with the decrease of temperature. A rapid decrease occurs in the chiral crossover region of $u, d$ quark, as shown in  Fig.~\ref{fig:5}.  It indicates that the value of speed of sound is sensitive to the change of dynamical quark mass. 
In the low-density region, a  minimum of $c^2_{s/\rho_B}$ appears near the deconfinement phase transition  for a given density~(chemical potential). A similar behavior exists in lattice QCD at zero chemical potential~\cite{Aoki06, Borsanyi14, Bazavov14, Philipsen13, Borsanyi20}. However, such a feature does not appear in the NJL model which cannot  describe the confinement-deconfinement phase transition~\cite{Ghosh06,Marty13,Deb16,Saha18}, which indicates that the color confinement also plays an important role on the speed of sound near the crossover phase transition line.

The value of $c^2_{s/\rho_B}$ is relatively smaller in the region of low temperature and density. 
The red dashed line is the boundary of vanishing sound velocity. Inside this boundary $c^2_{s/\rho_B}<0$, it means physically that $(\frac{\partial p }{\partial \epsilon})_{s/\rho_B}<0$. In this region, the mechanically stable condition is broken and the corresponding sound wave equation becomes a decay function. A perturbance can not be propagated like a sound wave in this situation. It can be seen that such a region lies in the interior of the unstable phase of the spinodal structure. Figure~\ref{fig:4} and \ref{fig:5}  also indicate that the speed of sound at the critical endpoint is small but not zero in the mean field approximation.

The value of $c^2_{s/\rho_B}$ reflects the speed of sound in an ideal fluid which can be approximately realized in heavy-ion collision experiments. On the other hand, $s/\rho_B$  is connected with the collision energy.
If the value of $c^2_{s/\rho_B}$ at a fixed energy can be extracted from the  charged particle multiplicity $\langle d N_{ch}/d\eta \rangle$~\cite{Gardim20,Sahu21,Biswas20}, we can access  the information of  phase transition using the relation between $c^2_{s/\rho_B}$ and QCD phase diagram. Furthermore, combining with the beam energy scan experiments, it  provides a possible way to diagnose the QCD phase structure. It  is also  inspiring for study on the gravitational signal from the cosmic QCD phase transition in which a constant speed of sound $\sqrt{1/3}$ is usually taken in literature.  

\subsection {Sound velocity at constant $\rho_B$ and $s$}

\begin{figure} [htbp]
\begin{minipage}{\columnwidth}
\centering
\includegraphics[scale=0.3]{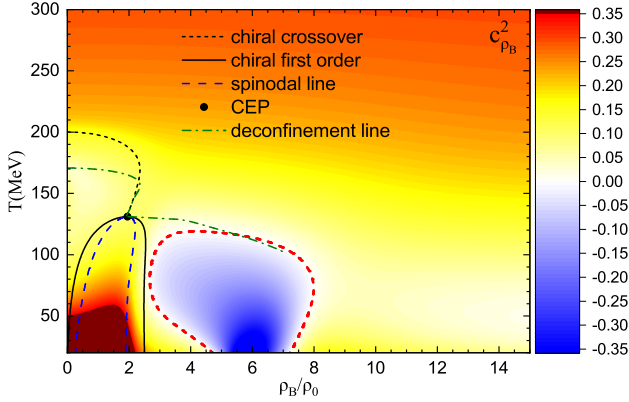}
\end{minipage}
\caption{   Contour map of $c^2_{\rho_B}$  in the $T-\rho_B$ plane. The red dashed line is the boundary of vanishing sound velocity.}
\label{fig:6}
\end{figure}

We present in Fig.~\ref{fig:6} the contour map of  $c_{\rho_B}^2$ in the $T-\rho_B$ panel. Besides at the high-temperature side, this figure shows that $c_{\rho_B}^2$ take relatively larger values in the region of low temperature and density. The value is even larger than $1/3$, in particular, in the metastable phase and unstable phase. There also exists a wide region~(inside the red line filled with the blue color) of $c_{\rho_B}^2<0$. More physically, it means that the $(\frac{\partial p}{\partial \epsilon})_{\rho_B}<0$ in this region, i.e., the pressure decreases with the increase of energy density along the line of constant density. The red line shows the boundary of vanishing sound speed at constant baryon density. 

\begin{figure} [htbp]
\begin{minipage}{\columnwidth}
\centering
\includegraphics[scale=0.3]{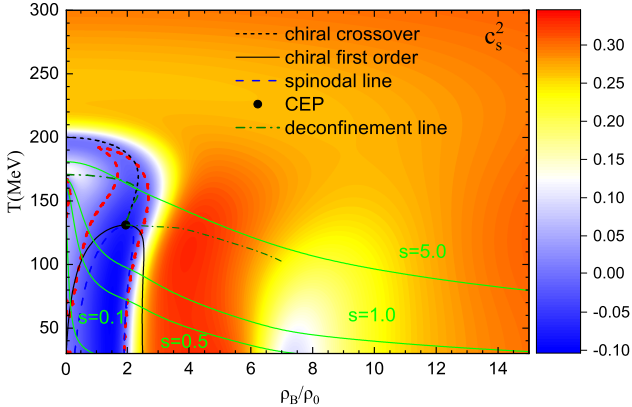}
\end{minipage}
\caption{ Contour map of $c^2_{s}$  in the $T-\rho_B$ plane. The red dashed line is the boundary of vanishing sound velocity. The green curves shows the paths of $s=0.1, 0.5, 1.0, 5$.}
\label{fig:7}
\end{figure}                    

The contour map of $c_s^2$ is presented in Fig.~\ref{fig:7}. The behavior of $c_s^2$ at low temperature and high density is similar with that of $c^2_{s/\rho_B}$, because the curves at constant $s$ and $s/\rho_B$ in the $T-\rho_B$ diagram are both roughly parallel with the density axis, as indicated in Fig.~\ref{fig:1}. and Fig.~\ref{fig:7}. However, the curves at constant $s$ at low density are almost perpendicular to those at constant $s/\rho_B$,  the resulting behaviors of $c_s^2$ and $c_{s/\rho_B}^2$ are quite different in the corresponding region. A distinct characteristic is the location of vanishing sound speed.   
The red dashed line in Fig.~\ref{fig:7} is the boundary of $c_{s}^2=0$. $(\frac{\partial p}{\partial \epsilon})_s$ takes minus values inside the boundary, which includes a wide range of the first-order phase transition and a region around the CEP. 

\begin{figure} [htbp]
\begin{minipage}{\columnwidth}
\centering
\includegraphics[scale=0.3]{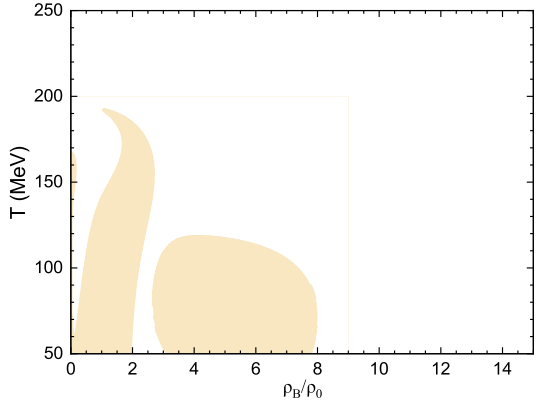}
\end{minipage}
\caption{ Regions of $(\frac{\partial \mu_B}{\partial T})_{s/\rho_B}<0$ in the $T-\rho_B$ panel. }
\label{fig:8}
\end{figure}
 
The region of $c_s^2<0$ in Fig.~\ref{fig:7}  and $c_{\rho_B}^2<0$ in Fig.~\ref{fig:6} are connected with the formula
\begin{equation}
\label{rhos}
\bigg(\frac{\partial \mu_B}{\partial T}\bigg)_{s/\rho_B}=\frac{\mu_B}{T}\frac{(\frac{\partial p}{\partial \epsilon})_{s}}{(\frac{\partial p}{\partial \epsilon})_{\rho_{B}}}=\frac{\mu_B}{T}\frac{c^2_s}{c^2_{\rho_B}}.
\end{equation}
When the condition $(\frac{\partial \mu_B}{\partial T})_{s/\rho_B}<0$ is fulfilled, one of the two physical quantities $c_s^2$ and $c_{\rho_B}^2$ takes a negative value. We show in Fig.~\ref{fig:8} the regions of $(\frac{\partial \mu_B}{\partial T})_{s/\rho_B}<0$. The numerical results indicate that there indeed exists the regions where $(\frac{\partial \mu_B}{\partial T})_{s/\rho_B}<0$ in the $T-\rho_B$ diagram.   
Comparing Fig.~\ref{fig:8} with the negative value regions in Fig.~\ref{fig:6} and Fig.~\ref{fig:7}, we can conclude that these numerical results confirm the formula in Eqs.~(\ref{rhos}).

The behaviors of $c_{\rho_B}^2$ and $c_s^2$ in the phase diagram can be used to study the fluid properties of quark gluon plasma. The values of  $c_{\rho_B}^2$ and $c_s^2$ are important parameters to indicate the intermediated process in the evolution of  a fluid. 
Eqs. (\ref{4}) and (\ref{5}) clearly show that $c_{\rho_B}^2$ and $c_s^2$  are connected with the temporal derivatives of temperature and  chemical potential, respectively.  Moreover, $c_{\rho_B}^2$ and $c_s^2$ are related to the bulk viscosity of a fluid. In particular they directly connect with the bulk viscosity coefficient. Exploring the relation between the bulk viscosity and phase transition is attractive to study the dissipation in the evolution of QGP. A further research  in this respect is undergoing.


\subsection {Sound velocity at constant $T$ and $\mu_B$}

\begin{figure} [htbp]
\begin{minipage}{\columnwidth}
\centering
\includegraphics[scale=0.3]{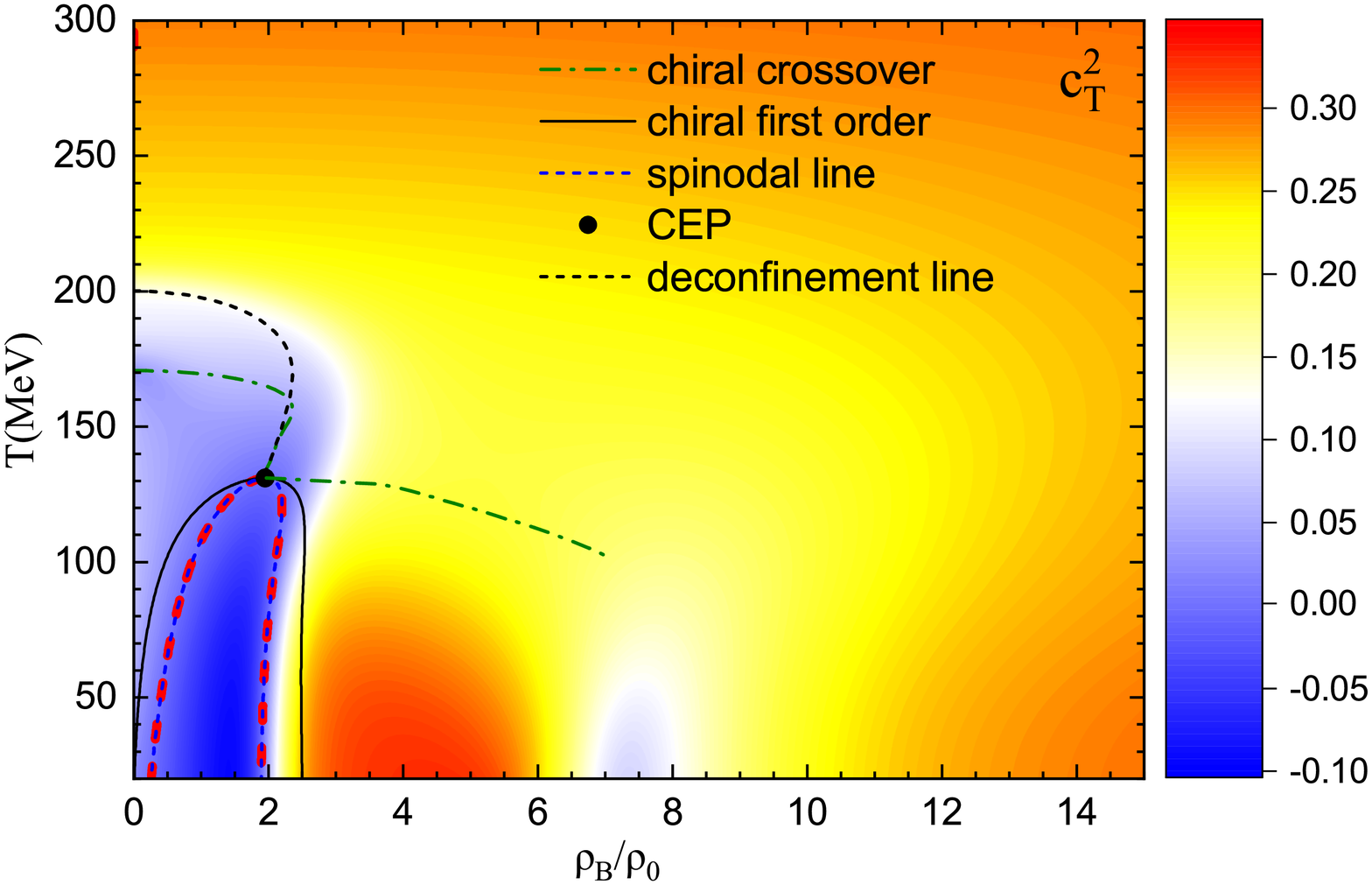}
\end{minipage}
\caption{Contour map of $c^2_{T}$  in the $T-\rho_B$ plane. The red dashed line is the boundary of vanishing sound velocity.}
\label{fig:9}
\end{figure}

\begin{figure} [htbp]
\begin{minipage}{\columnwidth}
\centering
\includegraphics[scale=0.3]{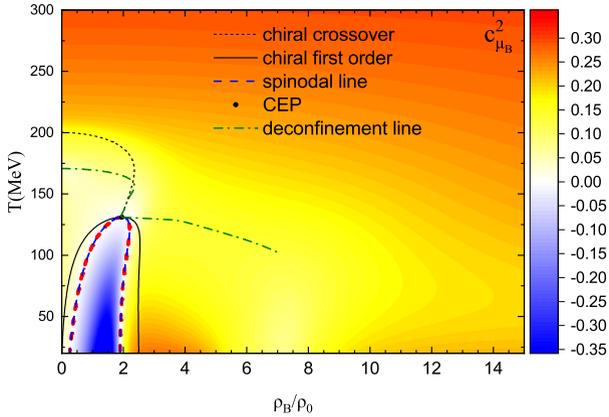}
\end{minipage}
\caption{ Contour map of $c^2_{\mu_B}$  in the $T-\rho_B$ plane. The red dashed line is the boundary of vanishing sound velocity.}
\label{fig:10}
\end{figure}

The contour maps of $c^2_T$ at constant temperature and $c^2_{\mu_B}$ at constant chemical potential in the $T-\rho_B$ panel are demonstrated in Fig.~\ref{fig:9} and Fig.~\ref{fig:10}, respectively.  The two figures show that $c^2_T$ and $c^2_{\mu_B}$ are both close to $1/3$ at high temperature.

The contour of $c^2_{T}$ looks in general like  that of $c^2_{s/\rho_B}$, because  the curves of constant $s/\rho_B$ in  the $T-\rho_B$ panel are almost parallel to the density axis in a wide range. The relative larger deviation lies in the range with densities smaller than the boundary of the first-order transition on the low-density side. The deviation produces different behaviors between $(\frac{\partial p}{\partial \epsilon})_T$ and $(\frac{\partial p}{\partial \epsilon})_{s/\rho_B}$ at low density. A crutial point is that the inflections points of $(\frac{\partial p}{\partial \epsilon})_T$ and $(\frac{\partial p}{\partial \epsilon})_{s/\rho_B}$  are different, i.e., the boundary of vanishing sound speed are different for the two cases.  


Both the figure~\ref{fig:9} and \ref{fig:10} show that the boundary of zero sound speed at constant temperature or chemical potential is just the spinodal line associated with the first-order phase transition.  However the  boundary of zero sound velocity at constant $s/\rho_B$ is in the interior of the spinodal structure as shown in Fig.~\ref{fig:4}. 
The negative values of $c^2_T$ and $c^2_{\mu_B}$ both appear in the unstable phase of the spinodal structure, since $(\frac{\partial p}{\partial \epsilon})_T$ and $(\frac{\partial p}{\partial \epsilon})_{\mu_B}$ are negative in this region.

From the behavior of $c^2_{T}$ derived above, we can approximately deduce the speed of sound in the quark core of a massive neutron star.  Since the richness of lepton including electron and muon approaches to zero at high density of a hybrid neutron star, and the richness of $u,d,s$ quark tends to be equivalent~\cite{Shao2013}, quite similar to the situation in this study at high density and low temperature. Therefore, it may be concluded to a certain degree that the squared speed of sound in the quark core of a massive neutron star gradually approaches to $1/3$ at high density. A further study on the speed of sound in neutron star matter with a hadron-quark phase transition will be conducted with the combination of observation data.

\section{Summary}   
In this work, we studied the speed of sound in QCD matter at finite temperature and density in the PNJL model.
We derived the behavior of speed of sound  under different definitions in the $T-\rho_B$ phase diagram including the stable phase,  metastable and unstable phases associated with the first-order phase transition.  We systematically discussed the relations between the speed of sound and QCD phase structure. 

The numerical results indicate that the squared speed of sounds under different definitions are all approaching to 1/3 at high temperature. 
However, the behaviors in the phase-transition region are closely related to the non-perturbative interaction and the phase structure. From the perspective of idea fluid evolution, more attention are put on the  speed of sound under isentropic condition. The calculation indicates that   $c^2_{s/\rho_B}$ is nonzero at the CEP under the mean field approximation, and the boundary of vanishing sound velocity is further derived.

We also obtained the contour maps of $c^2_{X}$ $(X=s,\rho_B, T,$ $ \mu_B)$, and analyzed their relations with the QCD phase transition, as well as the relations between different definitions. For each definition of speed of sound, we find that there exist one or several regions with $(\frac{\partial p}{\partial \epsilon})_X<0$, even in the stable phase for $X=s$ and $\rho_B$. 

The different definitions of speed of sound are involved in some important physics problems in nuclear physics, such as the evolution of quark gluon plasma, the cosmic QCD phase transition, the bulk viscosity of strongly interacting matter  and the equation of state of a hybrid neutron star.  The indepth investigations on these physics issues will be performed in the future.
%

\bmhead{Acknowledgments}
This work is supported by the National Natural Science Foundation of China under
Grant No. 11875213.

\bmhead{Data Availability} Data sharing not applicable to this article as no datasets were generated or analyzed during the current study.


\bibliography{sn-bibliography}

\begin{thebibliography}{99}
\bibitem{Song11} H. C. Song, S. A. Bass, U. Heinz, T. Hirano, and C. Shen, {\bf 106}, 192301 (2011).
\bibitem{Song112} H. C. Song, S. A. Bass, U. Heinz, Phys. Rev. C {\bf 83}, 024912 (2011).
\bibitem{Deb16}P. Deb, G. P. Kadam, and H. Mishra, Phys. Rev. D  {\bf 94}, 094002 (2016).
\bibitem{Gardim20} F. G. Gardim, G. Giacalone, M. Luzum, and J. Y.  Ollitrault, Nat. Phys.  {\bf 16,} 615 (2020).
\bibitem{Sahu21} D. Sahu, S. Tripathy, R. Sahoo, and A. R. Dash, Eur. Phys. J. A {\bf 56},  187  (2021). 
\bibitem{Biswas20} D. Biswas, K. Deka, A. Jaiswal, and S. Roy, Phys. Rev. C  {\bf 102}, 014912 (2020).

\bibitem{Sorensen21} A. Sorensen, D. Oliinychenko, V. Koch, and L. McLerran, Phys. Rev. Lett. {\bf 127},  042303  (2021).

\bibitem{Reed20} B. Reed and C. J. Horowitz, Phys. Rev. C {\bf 101}, 045803 (2020).
\bibitem{Kanakis20} A. Kanakis-Pegios, P. S. Koliogiannis, and Ch. C. Moustakidis, Phys. Rev. C {\bf 102}, 055801 (2020).
\bibitem{Han20} S. Han and M. Prakash, Astrophys. J.  {\bf 899}, 164 (2020).


\bibitem{Tews18} I. Tews, J. Carlson, S. Gandolfi, and S. Reddy,  Astrophys. J.  {\bf 860}, 149 (2018).
\bibitem{Greif19} S. K. Greif, G. Raaijmakers, K. Hebeler, A. Schwenk, and A. L. Watts, Mon. Not. R. Astron. Soc.  {\bf 485}, 5363 (2019).
\bibitem{Forbes19} M. M. Forbes, S. Bose, S. Reddy, D. Zhou, A. Mukherjee, and S. De,  Phys. Rev. D  {\bf 100}, 083010 (2019).
\bibitem{Drischler20} C. Drischler, S. Han, J. M. Lattimer, M. Prakash, S. Reddy, and T. Zhao, Phys. Rev. C {\bf 103}, 045808 (2021).
\bibitem{Essick20} R. Essick, I. Tews, P. Landry, S. Reddy, and D. E. Holz, Phys. Rev. C  {\bf 102}, 055803 (2020).
\bibitem{Han19} S. Han, M. A. A. Mamun, S. Lalit, C. Constantinou, and M. Prakash, Phys. Rev. D  {\bf 100}, 103022 (2019).
\bibitem{Kojo} T. Kojo,  AAPPS Bull. {\bf 31}, 11  (2021).
\bibitem{Jaikumar21} P. Jaikumar, A. Semposki,  M. Prakash, and C. Constantinou, Phys. Rev. D {\bf 103}, 123009 (2021).
\bibitem{Philipsen13} O. Philipsen, Prog. Part. Nucl. Phys. {\bf 70}, 55 (2013).
\bibitem{Borsanyi20} S. Bors\'anyi, Z. Fodor, J. N. Guenther, R. Kara, S. D. Katz, P. Parotto, A. Pasztor, C. Ratti, and K. K. Szabo, Phys. Rev. Lett. {\bf 125}, 052001 (2020).


\bibitem{Aoki06} Y. Aoki, G. Endrodi, Z. Fodor, S. D. Katz, K. K. Szabo, Nature (London)    {\bf 443}, 675 (2006).
\bibitem{Borsanyi14} S. Bors\'anyi, Z. Fodor, C. Hoelbling, S. D. Katz, S. Krieg,
and K. K. Sabz\'o, Phys. Lett. B {\bf 730},  99 (2014).
\bibitem{Bazavov14} A. Bazavov {\it et al.} (hotQCD Collaboration), Phys. Rev. D. {\bf 90}, 094503  (2014). 
\bibitem{Motta18}M. Motta, R. Stiele, W. M. Alberico, A. Beraudo, Eur. Phys. J. C {\bf 80},  770  (2020). 
\bibitem{Ghosh06} S. K. Ghosh, T. K. Mukherjee, M. G. Mustafa, and R. Ray, Phys. Rev. D  {\bf 73}, 114007 (2006). 
\bibitem{Marty13} R. Marty, E.  Bratkovskaya, W. Cassing, J. Aichelin, and H. Berrehrah, Phys. Rev. C  {\bf 88}, 045204 (2013).
\bibitem{Saha18} K. Saha, S. Ghosh, S. Upadhaya, S. Maity, Phys. Rev. D  {\bf 97}, 116020 (2018).
\bibitem{zhao20} Y. P. Zhao, Phys. Rev. D  {\bf 101}, 096006, (2020).
\bibitem{Schaefer10} B. J. Schaefer, M. Wagner, and J. Wambach, Phys. Rev. D {\bf 81}, 074013 (2010).
\bibitem{Abhishek18} A. Abhishek, H. Mishra, and S. Ghosh, Phys. Rev. D  {\bf 97}, 014005 (2018).

\bibitem{Venugopalan92} R. Venugopalan and M. Prakash, Nucl. Phys.  {\bf A546,} 718 (1992).
\bibitem{Bluhm14} M. Bluhm, P. Alba, W. Alberico, A. Beraudo, and C. Ratti, Nucl. Phys.  {\bf A929}, 157 (2014).
\bibitem{Khaidukov18} Z. V. Khaidukov, M. S. Lukashov, and Yu. A. Simonov, Phys. Rev. D  {\bf 98}, 074031 (2018).
\bibitem{Khaidukov19} Z. V. Khaidukov and Yu. A. Simonov, Phys. Rev. D  {\bf 100}, 076009 (2019).
\bibitem{Mykhaylova21}V. Mykhaylova and C. Sasaki, Phys. Rev. D  {\bf 103}, 014007 (2021).
\bibitem{he2022} W. B. He, G. Y. Shao, X. Y. Gao, X. R. Yang, and C. L. Xie, Phys. Rev. D  {\bf 105}, 094024 (2022).
\bibitem{Mishustin1999} I. N. Mishustin, Phys. Rev. Lett. {\bf 82},  4779  (1999).
\bibitem{Randrup2004} J. Randrup, Phys. Rev. Lett. {\bf 92},  122301  (2004).
\bibitem{Koch2005} V. Koch, A. Majumder, and J. Randrup, Phys. Rev. C {\bf 72},  064903  (2005).
\bibitem{Sasaki2007} C. Sasaki, B. Friman, and K. Redlich, Phys. Rev. Lett. {\bf 99},  232301  (2007).
\bibitem{Sasaki2008} C. Sasaki, B. Friman, and K. Redlich, Phys. Rev. D {\bf 77},  034024  (2008).
\bibitem{Randrup2009} J. Randrup, Phys. Rev. C {\bf 79},  054911  (2009); Phys. Rev. C {\bf 82},  034902  (2010).
\bibitem{Steinheimer2012} J. Steinheimer and J. Randrup, Phys. Rev. Lett. {\bf 109},  212301  (2012).  
\bibitem{Steinheimer2016} J. Steinheimer and J. Randrup, Eur. Phys. J. A {\bf 52},  239  (2016).  
\bibitem{Li2016} F. Li and C. M. Ko, Phys. Rev. C {\bf 93},  035205  (2016). 
\bibitem{Steinheimer2017} J. Steinheimer and V. Koch, Phys. Rev. C. {\bf 96},  034907  (2017). 
\bibitem{Shao2020} G. Y. Shao, X. Y. Gao, and W. B. He, Eur. Phys. J. A {\bf 56}, 115 (2020).
\bibitem{Ahmadvand2018} M. Ahmadvand, K. Bitaghsir Fadafan, Phys. Lett. B {\bf 779}, 1 (2018).
\bibitem{Tenkanen2022} Tuomas V. I. Tenkanen and Jorinde van de Vis, J. High Energ. Phys. {\bf 2022}, 302 (2022). 
\bibitem{Robner07} S. R\"{o}{\ss}ner, C. Ratti, and W. Weise, Phys. Rev. D {\bf 75},  034007 (2007).
\bibitem{Rehberg96} P. Rehberg, S. P. Klevansky, and J. H\"ufner, Phys. Rev. C {\bf 53},  410 (1996).
\bibitem{Costa10}  P. Costa, M. C. Ruivo, C. A. de Sousa, and H. Hansen, Symmetry {\bf 2}, 1338 (2010).
\bibitem{Shao2018} G. Y. Shao, Z. D. Tang, X. Y. Gao, and W. B. He, Eur. Phys. J . C {\bf 78}, 138 (2018).
\bibitem{Chomaz2004} P. Chomaz, M. Colonna, and J. Randrup, Phys. Rep. {\bf 389} 263 (2004).
\bibitem{Shao20202} G. Y. Shao, X. Y. Gao, and W. B. He, Phys.Rev.D {\bf 101}, 074029 (2020).
\bibitem{Shao2013} G. Y. Shao, M. Colonna, M. Di Toro,  Y. X.  Liu, and B. Liu, Phys.Rev.D {\bf 87}, 096012 (2013).
\end{thebibliography}


\newpage
\appendix 
\section{Derivations of the formulae of speed of sound under different definitions in the temperature and density space}
\label{sec:appd}
The general definition of speed of sound is 
\begin{equation}
c_{X}^2=\left(\frac{\partial p}{\partial \epsilon}\right)_{X},
\end{equation}
where $X$ is a physics quantum fixed in the calculation of sound speed. 
In practice, the squared speed of sound
$c^2_{X}$ $(X=s/\rho_B, s,\rho_B, T, \mu_B)$  under different conditions are taken in dealing with different physics problems. With the basic definitions of speed of sound, calculation  can only be done along some special paths.

To calculate the speed of sound under different definitions in the whole $T-\rho_B$ space, it is necessary to derive the corresponding formulae  in terms of temperature and density. Using the Jacobian formula in thermodynamics, we can derive
\begin{eqnarray}
c_{X}^{2}\left(T, \rho_{B}\right)&=&\left(\frac{\partial p}{\partial \epsilon}\right)_{X}=\frac{\partial(p, X)}{\partial(\epsilon, X)}=\frac{\frac{\partial(p, X)}{\partial\left(T, \rho_{B}\right)}}{\frac{\partial(\epsilon, X)}{\partial\left(T, \rho_{B}\right)}}  \nonumber \\
& =&\frac{\left(\frac{\partial p}{\partial T}\right)_{\rho_{B}}\left(\frac{\partial X}{\partial \rho_{B}}\right)_{T}-\left(\frac{\partial p}{\partial \rho_{B}}\right)_{T}\left(\frac{\partial X}{\partial T}\right)_{\rho_{B}}}{\left(\frac{\partial \epsilon}{\partial T}\right)_{\rho_{B}}\left(\frac{\partial X}{\partial \rho_{B}}\right)_{T}-\left(\frac{\partial \epsilon}{\partial \rho_{B}}\right)_{T}\left(\frac{\partial X}{\partial T}\right)_{\rho_{B}}}
\end{eqnarray}

%

According to the thermodynamic characteristic function in the giant canonical ensemble, it is convenient to get the following relations for isospin symmetric matter
\begin{equation}
\left(\frac{\partial p}{\partial T}\right)_{\rho_{B}}=s+\rho_{B}\left(\frac{\partial \mu_{B}}{\partial T}\right)_{\rho_{B}},
\end{equation}
\begin{equation}
\left(\frac{\partial p}{\partial \rho_{B}}\right)_{T}=\rho_{B}\left(\frac{\partial \mu_{B}}{\partial \rho_{B}}\right)_{T},
\end{equation}
\begin{equation}
\left(\frac{\partial \epsilon}{\partial T}\right)_{\rho_{B}}=T\left(\frac{\partial s}{\partial T}\right)_{\rho_{B}},
\end{equation}
\begin{equation}
\left(\frac{\partial \epsilon}{\partial \rho_{B}}\right)_{T}=T\left(\frac{\partial s}{\partial \rho_{B}}\right)_{T}+\mu_{B},
\end{equation}
\begin{equation}
\left(\frac{\partial {(s/\rho_B)}}{\partial T}\right)_{\rho_B}=\frac{1}{\rho_{B}}\left(\frac{\partial s}{\partial T}\right)_{\rho_{B}},
\end{equation}
and
\begin{equation}
\left(\frac{\partial{ (s/\rho_B)}}{\partial \rho_{B}}\right)_{T}=\frac{1}{\rho_{B}}\left(\frac{\partial s}{\partial \rho_{B}}\right)_{T}-\frac{s}{\rho_{B}^{2}}
\end{equation}

For the different constraint conditions, $X=s/\rho_B, s,\rho_B, T, \mu_B$, we can derived the corresponding formulae of speed of sound  as follows

\begin{equation}\label{}
\!c^2_{s/\rho_B}\!=\!\frac{\!s^{2}\!+\!\rho_{B}^{2}\!\left[\!\left(\frac{\!\partial \mu_{B}}{\!\partial \rho_{B}}\!\right)_{\!T}\!\left(\frac{\partial s}{\!\partial T}\!\right)_{\!\rho_{\!B}}\!-\!\left(\frac{\partial \mu_{\!B}}{\partial T}\!\right)_{\!\rho_{\!B}}\!\left(\frac{\partial s}{\partial \rho_{\!B}}\!\right)_{\!T}\!\right]\!+\!s \rho_{\!B}\!\left[\!\left(\frac{\partial \mu_{\!B}}{\partial T}\!\right)_{\!\rho_{\!B}}\!-\!\left(\frac{\partial s}{\partial \rho_{\!B}}\!\right)_{\!T}\right]}{\left(T s+\mu_{B} \rho_{B}\right)\left(\frac{\partial s}{\partial T}\right)_{\rho_{B}}},
\end{equation}
\begin{equation}
c_{s}^{2}=\frac{\rho_{B}\left[\left(\frac{\partial s}{\partial T}\right)_{\rho_{B}}\left(\frac{\partial \mu_{B}}{\partial \rho_{B}}\right)_{T}-\left(\frac{\partial s}{\partial \rho_{B}}\right)_{T}\left(\frac{\partial \mu_{B}}{\partial T}\right)_{\rho_{B}}\right]-s\left(\frac{\partial s}{\partial \rho_{B}}\right)_{T}}{\mu_{B}\left(\frac{\partial s}{\partial T}\right)_{\rho_{B}}},
\end{equation}
\begin{equation}
c_{\rho_{B}}^{2}=\frac{s+\rho_{B}\left(\frac{\partial \mu_{B}}{\partial T}\right)_{\rho_{B}}}{T\left(\frac{\partial s}{\partial T}\right)_{\rho_{B}}},
\end{equation}
\begin{equation}
c_{T}^{2}=\frac{\rho_{B}\left(\frac{\partial \mu_{B}}{\partial \rho_{B}}\right)_{T}}{T\left(\frac{\partial s}{\partial \rho_{B}}\right)_{T}+\mu_{B}},
\end{equation}
and
\begin{equation}\label{}
c_{\mu_{B}}^{2}=\frac{s\left(\frac{\partial \mu_{B}}{\partial \rho_{B}}\right)_{T}}{T\left[\left(\frac{\partial s}{\partial T}\right)_{\rho_{B}}\left(\frac{\partial \mu_{B}}{\partial \rho_{B}}\right)_{T}-\left(\frac{\partial \mu_{B}}{\partial T}\right)_{\rho_{B}}\left(\frac{\partial s}{\partial \rho_{B}}\right)_{T}\right]-\mu_{B}\left(\frac{\partial \mu_{B}}{\partial T}\right)_{\rho_{B}}}.
\end{equation}

\end{document}